\documentclass{jfm}
\usepackage{color}
\usepackage{amstext}
\usepackage{amssymb}
\usepackage{graphicx}
\usepackage{amsmath}
\usepackage{tcolorbox}

\newcommand\const{\mathrm{const}}

\newcommand\vX{\boldsymbol{X}}

\newcommand\vF{\boldsymbol{F}}

\newcommand\va{\boldsymbol{a}}
\newcommand\vb{\boldsymbol{b}}

\newcommand\vf{\boldsymbol{f}}
\newcommand\vh{\boldsymbol{h}}

\newcommand\vx{\boldsymbol{x}}

\begin{document}

{\title[ Distinguished Limits and Vibrogenic Force] {Distinguished Limits and Vibrogenic Force\\ revealed by Newton's Equation\\ with Oscillating Force}}

\author[V. A. Vladimirov]
{V.\ns A.\ns V\ls l\ls a\ls d\ls i\ls m\ls i\ls r\ls o\ls v}


\affiliation{York University, UK; Sultan Qaboos University, Oman; Leeds University, UK}

\pubyear{2014} \volume{xx} \pagerange{xx-xx}

\setcounter{page}{1}\maketitle \thispagestyle{empty}

\begin{abstract}

In this paper, we analyse the basic ideas of Vibrodynamics and the two-timing method.
To make our analysis most instructive, we have chosen the Newton's equation with a general oscillating force.
We deal with its asymptotic solutions in the high frequency limit.
Our treatment is simple but general.
The targets of our study are \emph{the distinguished limits} and \emph{the universal  vibrogenic force}.
The aim of \emph{the distinguished limit procedure} is to identify how the small parameter can appear in an equation.
The proper appearance of a small parameter leads to \emph{valid successive approximations}, and, in particular, to closed systems of averaged equations.
We show, that there are only two distinguished limits.
This means that Newton's equation, with high-frequency forcing, has  two types of interesting asymptotic solutions.
The key item in the averaged equations for all distinguished limits is \emph{the unique vibrogenic force}.
The current state-of-the-art in this area is: a large number of particular examples are well-known, effective and advanced general methods (like the Krylov-Bogolyubov approach) are well developed.
However, the presented general and simple analysis of distinguished limits and the vibrogenic force, formulated as a compact practical guide, is novel.
An advantage of our treatment is the possibility of its straightforward use for various ODEs and PDEs with oscillating coefficients.
\end{abstract}

\section{Introduction \label{sect01}}

This paper is devoted to the analysing of the basic ideas of Vibrodynamics and the two-timing method.
To do this most instructive,  we use Newton's equation with a general high-frequency oscillating force.
Under oscillations we understand any time-periodic process, including rotation.
Our attention is focused at \emph{the distinguished limits (DLs)} and \emph{ the vibrogenic force}.
There are different definitions of \emph{DL}, see \emph{e.g.} \cite{Nayfeh, Kevorkian}.
All these definitions operate with  `proper' relations between different terms in a considered equation, such  relations must lead to valid mathematical results.
For example, if we have two independent small parameters $\varepsilon_1$ and $\varepsilon_2$ in an equation, then we cannot say, what is greater $\varepsilon^2_1$ or $\varepsilon_2$. 
As the result, the successive approximations cannot be derived.
For resolving this problem, we should study different pathes in the $(\varepsilon_1,\varepsilon_2)$-plane, such that $(\varepsilon_1,\varepsilon_2)\to (0,0)$. 
Such pathes can be presented in a parametric form $\varepsilon_1(\varepsilon)$, $\varepsilon_2(\varepsilon)$ with the only small parameter $\varepsilon\to 0$. 
If such a limit brings valid asymptotic results, then it is called \emph{a distinguished limit (DL)}.
A common point of confusion is: \emph{DLs} are usually used as \emph{know-how}, without any comments to their appearance and to the presence or absence of any alternatives.
Often, \emph{DL} is implicitly used without naming it.
The existence of \emph{DLs} demonstrates that equations are `smart' by themselves: the structure of an equation dictates the placing a small parameter within it.
It is important to emphasise that the \emph{DL procedure} represents \emph{a purely mathematical tool}.
The fitting of a particular physical problem into the \emph{DLs} represents the next task.
The term \emph{vibrogenic force}, for the averaged force, generated by oscillations/vibrations was introduced by \cite{Yudovich}.
Similarly,  any averaged product of oscillating functions can be called a \emph{vibrogenic term}.
A number of references could be very significant, the directly related treatments are \cite{Vladimirov, Vladimirov1, Yudovich}.

In this paper we show that there are two \emph{DLs} for Newton's equation.
Our results are exposed briefly, but with a natural generality and simplicity.
In Section 2, we present the general setting of the two-timing method and Vibrodynamics.
All the required notations and suggestions are introduced in Section 3.1. The next
Section 3.2 is devoted to the general properties of considered distinguished limits.
The list of averaged equations is given and analysed in Section 3.3.
In Section 4, we illustrate our approach  by the paradigm example of the Stephenson-Kapitza pendulum (a pendulum with an oscillating pivot), where our attention is focused at fitting the pendulum's equation into the described \emph{DLs}.
Section 5 contains the formulation of the practical algorithm for using our results.
Finally, the detailed calculations of successive approximations are presented in the Appendix.

\section{Two-Timing Setting}

We consider a system of ODEs
\begin{equation}\label{ODE}
 d^2\vx^\ast/dt^{*2}=\vf^\ast(\vx^\ast,\tau),\qquad \tau\equiv\omega^\ast t^\ast
\end{equation}
where $\vx^\ast=(x_1^\ast,x_2^\ast,x_3^\ast)$ and $t^\ast$ are cartesian coordinates and time, $\vx^\ast(t^\ast)$ is the unknown function, $\vf^\ast(\vx^\ast,\tau)$ represents a given oscillating force, a constant mass of a particle is included in the definition of force, $\omega^\ast$ is a given frequency, and asterisks mark the dimensional variables.
We assume that there are mutually independent characteristic scales of length $L$, time $T$, and force $F$. Then, the above equation in the dimensionless form is:

\begin{equation}\label{ODE1}
 \boxed{\frac{d^2\vx}{dt^2}=(FT^2/L)\vf(\vx,\tau),\qquad}
\end{equation}
where
\begin{equation}\nonumber\label{variables}
 \vx^\ast=L\vx,\quad t^\ast =T t, \quad  \vf^\ast=F\vf,\quad  \omega^\ast=\omega/T, \quad\tau=\omega t=\omega^\ast t^\ast
\end{equation}
The first step of the two-timing method is the introduction of two (\emph{originally mutually dependent}) time-variables $s$ and $\tau$; we call $s$ \emph{slow time} and $\tau$ \emph{fast time}:
\begin{eqnarray}
&&\tau\equiv\omega {t},\quad s\equiv t/\omega^\alpha;\quad  \alpha=\const>-1 ,\quad \omega\gg 1
\label{exact-2}
\end{eqnarray}
The restriction $\alpha>-1$ establishes the roles of $\tau$ and $s$ as \emph{the fast time variable} and \emph{the slow time variable} for $\omega\to \infty$.
The use of the chain rule brings \eqref{ODE1} to the form
\begin{eqnarray}
&&\left(\omega\frac{\partial}{\partial\tau}+\frac{1}{\omega^{\alpha}}\frac{\partial}{\partial s}\right)^2\vx =(FT^2/L)\vf(\vx,\tau)\label{PDE}
\end{eqnarray}
where we have two independent scaling parameters $\omega$ and $(FT^2/L)$.
Next, we reduce the problem to a single parameter by introducing a functional dependence $FT^2/L\equiv \omega^\beta$, where $\beta=\const$.
Then \eqref{PDE} is transformed to the equation
\begin{eqnarray}
&&\boxed{{\vx}_{\tau\tau}+ 2\varepsilon\vx_{\tau s}+\varepsilon^2 \vx_{ss} =\varepsilon^\kappa\vf(\vx,\tau),
\quad \varepsilon\equiv \omega^{-1-\alpha},\quad \kappa\equiv\frac{2-\beta}{1+\alpha}}\label{PDE1}
\end{eqnarray}
where $\varepsilon$ is the only small parameter and the subscripts $\tau$ and $s$ stand for the partial derivatives.
The key suggestion of the two-timing method is
\begin{eqnarray}
&&\boxed{\emph{$\tau$ and $s$ are treated as mutually independent variables}}\label{key}
\end{eqnarray}
 which converts  \eqref{PDE1} from an ODE with the only independent variable $t$ into a PDE with two independent variables $\tau, \, s$.
 As a result, any solutions of \eqref{PDE1} must have the functional form:
\begin{eqnarray}
&& \boxed{\vx={\vx}(s, \tau)}\label{soln}
\end{eqnarray}
A regular perturbation procedure starts from substituting the series
\begin{eqnarray}\label{series}
\vx(s,\tau)=\sum_{n=0}^\infty \varepsilon^n \vx_n(s, \tau),\quad n=0,1,2,\dots
\end{eqnarray}
into \eqref{PDE1}.
After ${\vx}(s, \tau)$  is found, one can use \eqref{exact-2} to return to the original variable $t$: $\vX(t)\equiv{\vx}(s(t), \tau(t))$.
Mathematical justification of the obtained solution $\vX(t)$ consists of proving that the difference between $\vX(t)$ and the exact solution to \eqref{ODE1} is small.
 In this paper, we accept \eqref{key} and \eqref{soln} and do not consider any mathematical justifications.

\section{Asymptotic Procedure and Distinguished Limits}

\subsection{\underline{Required Notations, Operations, and Suggestions}}

To make our treatment self-consistent and efficient, we introduce some notations and agreements.
Let $g=g(s,\tau)$  be \emph{any dimensionless function, which could be scalar, vectorial, or tensorial}.  We accept that:

\textbf{(A)}  \emph{The order agreement:} $g\sim {O}(1)$, and its $s$-, and $\tau$-derivatives (required for our consideration) are also ${O}(1)$.
All small parameters are represented by various degrees of $\varepsilon$ only, they appear as explicit multipliers.
If we do not accept this, then the use of the successive approximations is impossible.
In particular, it makes impossible the change of $\kappa$ by dividing \eqref{PDE1} by any degree of $\varepsilon$.

\textbf{(B)} $\tau$\emph{-periodicity:} $g$  is $2\pi$-periodic in $\tau$, \emph{i.e.}\ $g(s, \tau)=g(s, \tau+2\pi)$, which represents a technical simplification;

\textbf{(C)} \emph{Average, bar-functions and tilde-functions:} $g$ has an average given by
\begin{equation}\label{aver}
\overline{g}\equiv \langle {g}\,\rangle \equiv \frac{1}{2\pi}\int_{\tau_0}^{\tau_0+2\pi}
g(s, \tau)\, d \tau \qquad \forall\ \tau_0=\const,
\end{equation}
hence any $g(s,\tau)$ can be split into the sum of its average part and purely oscillating part $g(s, \tau)=\overline{g}(s)+\widetilde{g}(s, \tau)$,
where  the \emph{tilde-function} (or  purely oscillating function) is such that $\langle \widetilde g\, \rangle =0$ and the \emph{bar-function} $\overline{g}(s)$ is $\tau$-independent. The introduced functions $\vx(s,\tau)$, $\vf(\vx,t)$, and $\vx_n(s,t)$ represent the combinations of bar- and tilde- parts, \emph{e.g.} $\vx_n(s,\tau)=\overline{\vx}_n(s)+\widetilde{\vx}_n(s,\tau)$;

\textbf{(D)} \emph{Tilde-integration:} $\widetilde{g}^{\tau}$ (with a superscript $\tau$) stands for the operation:
\begin{equation}\nonumber
\widetilde{g}^\tau\equiv G-\overline{G},\quad G(s,\tau)\equiv\int_0^\tau \widetilde{g}(s,\tau')\, d \tau'
\label{ti-integr0}
\end{equation}
We call it \emph{the tilde-integration}, since it keeps the result within the tilde-class.
The tilde-integration is
inverse to the $\tau$-differentiation $({{\widetilde{g}}}^{\tau})_{\tau}=({g}_{\tau})^{\tau}={\widetilde{g}}$;

\textbf{(E)} \emph{Taylor series:}
The related to \eqref{series}   expansion for $\vf=(f_{1},f_{2},f_{3})$ appears as
\begin{eqnarray}\label{Taylor}
\vf(\vx,\tau)=\vf_0+\varepsilon\,  (\vx_{1}\cdot\nabla_0) \vf_{0,k}
+\varepsilon^2 (\vx_{2}\cdot \nabla_{0})\vf_{0} +\frac{1}{2} x_{1k}x_{1l} \vf_{0,kl}+O(\varepsilon^3),
\end{eqnarray}
We use the summation convention and shorthands $\vf_{0}\equiv \vf (\vx_0,\tau)$, $\vf_{0,k}\equiv {\partial \vf_{0}}/{\partial x_{0k}}$, $\vf_{0,kl}\equiv {\partial^2 \vf_{0}}/{\partial x_{0k}\partial x_{0l}}$, $\vx_0=(x_{01},x_{02},x_{03})$,
$\nabla_0\equiv(\partial/\partial x_{01},\partial/\partial x_{02},\partial/\partial x_{03})$,  $\vf_0=(f_{01},f_{02},f_{03})$, \emph{etc.}
For brevity, we simulaneously use both the vectorial notations and the component/subscript notations.

\textbf{(F)} \emph{The class of motion:} We accept that
\begin{eqnarray}\label{smal-tilde}
\vx_0(s,t)=\overline{\vx}_0(s)\quad \text{or}\quad \widetilde{\vx}_0(s,t)\equiv 0
\end{eqnarray}
\emph{The constraint \eqref{smal-tilde} means that the amplitude of oscillations/vibrations is small compared with the amplitude of the average motion.}
This \emph{constraint} is both mathematical and physical.
\emph{Mathematically:} If $\widetilde{\vx}_0\neq 0$, then the series \eqref{Taylor} has the coefficients like $\vf_0\equiv\vf(\overline{\vx}_0+\widetilde{\vx}_0, \tau)$, \emph{etc.}
Such an expression, for a general function $\vf$, makes impossible the split of \eqref{Taylor} into their bar parts and tilde parts,
and hence the procedure of successive approximations fails.
\emph{Physically:} If $\widetilde{\vx}_0\neq 0$, then the  main term of velocity grows to infinity  $|\omega \widetilde{\vx}_{0\tau}|\to \infty$ as $\omega\to \infty$. Then, $\widetilde{\vx}_{0\tau}=0$ represents a natural physical constraint, while $\widetilde{\vx}_{0}=0$ follows after its integration, due to the zero average of a tilde-function.
It is important, that we can consider only the versions of the equation \eqref{PDE1} not contradicting to \eqref{smal-tilde}.

\subsection{\underline{General Properties of Distinguished Limits} }

The consistency between  equations \eqref{PDE1}, \eqref{series} and the agreement \textbf{(A)} above, require that parameter $\kappa$ in  \eqref{PDE1} represents a non-negative integer: $\kappa=0,1,2,3,\dots$.
\emph{Then each DL appears as the value of $\kappa$ which, after the substitution of \eqref{series} into \eqref{PDE1}, leads to a valid procedure of successive approximations}.
To obtain a full set of asymptotic equations, one should consider all available values of $\kappa$.
Our calculations for \eqref{PDE1}-\eqref{Taylor} represent the standard procedure of successive approximations, complimented by splitting solutions into the averaged part and the purely oscillating part.
This procedure allows to calculate the full solution, including its average and oscillating parts.
However, for practical use, we will present only the calculations until the approximation, in which the \emph{vibrogenic force} $\overline{\vF}^{\text{vg}}$ appears.
Below we present the final results for \emph{DLs}, while all the calculations of successive approximations are moved to the Appendix.
The general results for different $\kappa$ are:

$\bullet$ $\kappa=0$: The amplitudes of the oscillatory part and the averaged part in $\vx_0(s,\tau)=\overline{\vx}_0(s)+\widetilde{\vx}_0(s,\tau)$ are of the same order.
It violates our main constraint \eqref{smal-tilde}, and therefore it is out of our consideration.

$\bullet$  $\kappa=1$: A valid/closed system of successive approximations does exist, we refer to it as \emph{{Distinguished Limit-1}} or \emph{{DL-1}}.
   $\overline{\vF}^{\text{vg}}$ appears in the zeroth approximation of the averaged motion.

$\bullet$ $\kappa=2$: A valid/closed system does exist, we refer to it as \emph{{DL-2}}.
The same as in \emph{DL-1} vibrogenic force $\overline{\vF}^{\text{vg}}$ appears in the second approximation of the averaged motion.

$\bullet$ $\kappa=3$: A closed system of successive approximations  appears, however it degenerates.
It appears since the forcing is so small that $\overline{\vx}_0(s)$ represents a free rectilinear motion with a constant velocity and 
$\overline{\vF}^{\text{vg}}$  appears only in the fourth approximation of the averaged motion.

$\bullet$ $\kappa=N> 3$, with an integer $N$: Closed systems also appears, however the degrees of their degeneration are even higher than for $\kappa=3$.
The orders of approximation, describing the rectilinear motion and the the orders of averaged equations where $\overline{\vF}^{\text{vg}}$ appears, are increasing with the grow of $N$.


\subsection{\underline{The List of Averaged Equations for DL-1,2,3}}

In this subsection, we present only the averaged equations and discuss their properties.
The related calculations of successive approximations are given in the Appendix.

\vskip 3mm
\noindent\fbox{%
    \parbox{\textwidth}{%
    The averaged \emph{DL-1} equations have been obtained by considering three successive approximations $n=0,1,2$ of \eqref{PDE1} for $\kappa=1$:
\begin{eqnarray}
&&\overline{\vx}_{0ss}=\overline{\vF}^{\text{vg}},\quad \text{the required constraint}\ \overline{\vf}_0\equiv 0\label{1-bar-0}
\end{eqnarray}
}}
\vskip 3mm
\noindent\fbox{%
    \parbox{\textwidth}{%
    The averaged \emph{DL-2} equations have been obtained by considering five successive approximations $n=0,1,2,3,4$ of \eqref{PDE1} for $\kappa=2$:
\begin{eqnarray}
&&\overline{\vx}_{0ss}=\overline{\vf}_0,\label{2-bar-0}\\
&&\overline{\vx}_{1ss}=(\overline{\vx}_1\cdot\nabla_0)\overline{\vf}_0\label{2-bar-1}\\
&&\overline{\vx}_{2ss}=(\overline{\vx}_2\cdot\nabla_0)\overline{\vf}_0+
\frac{1}{2}\overline{x}_{1i}\overline{x}_{1k} \overline{\vf}_{0,ik}+\overline{\vF}^{\text{vg}}
\label{2-bar-2}
\end{eqnarray}
}}

\vskip 3mm
\noindent\fbox{%
    \parbox{\textwidth}{%
   The averaged $\kappa=3$ equations have been obtained by considering seven successive approximations $n=0,1,2,3,4,5,6$ of \eqref{PDE1}:
\begin{eqnarray}
&&\overline{\vx}_{0ss}=0\quad \text{or}\quad\overline{\vx}_0=\va s+\vb,\ \text{where}\ \va\ \text{and}\  \vb\ \text{are arbitrary constants}\label{3-bar-0}  \\
&&\overline{\vx}_{1ss}=\overline{\vf}_0\label{3-bar-1}\\
&&\overline{\vx}_{2ss}=(\overline{\vx}_1\cdot\nabla_0)\overline{\vf}_0\label{3-bar-2}\\
&&\overline{\vx}_{3ss}=(\overline{\vx}_2\cdot\nabla_0)\overline{\vf}_0+\overline{x}_{1i}\overline{x}_{1k}\overline{\vf}_{0,ik}/2\label{3-bar-3}\\
&&\overline{\vx}_{4ss}=(\overline{\vx}_3\cdot\nabla_0)\overline{\vf}_0+\overline{x}_{1i}\overline{x}_{2k}\overline{\vf}_{0,ik}+
\overline{x}_{1i}\overline{x}_{1k}\overline{x}_{1l}\overline{\vf}_{0,ikl}/6+\overline{\vF}^{\text{vg}}\label{3-bar-4}
\end{eqnarray}
}}
\vskip 3mm
\noindent\fbox{%
    \parbox{\textwidth}{%
 In all cases $\kappa=1,2,3$, the expression for vibrogenic force is the same:
\begin{eqnarray}
&&\quad\overline{\vF}^{\text{vg}}\equiv -\langle(\widetilde{\vf}_{0}^\tau\cdot\nabla_0)\widetilde{\vf}_{0}^\tau\rangle
\label{F-vibr}
\end{eqnarray}
}}\vskip 3mm \noindent
The important properties of the above averaged equations with $\kappa=1,2,3$ are:
\vskip 1mm \noindent
\textbf{(i)} $\kappa=1,2,3$:  All three systems are presented only until the approximation, where $\overline{\vF}^{\text{vg}}$ appears;
\vskip 1mm \noindent
\textbf{(ii)}  $\kappa=1,2,3$:  For $\overline{\vF}^{\text{vg}}\equiv 0$, each averaged system coincides with the successive approximations of the original equation \eqref{ODE1}, taken for the $\tau$-independent function $\vf$;
\vskip 1mm \noindent
\textbf{(iii)}  \emph{DL-1}: Equation \eqref{1-bar-0} provides the promising option of entering $\overline{\vF}^{\text{vg}}$ in the zero/main approximation.
 The constraint $\overline{\vf}_0\equiv 0$ restricts the applications to the purely vibrational (in zero approximation) force $\vf_0=\widetilde{\vf}_0\neq 0$. 
 Such a force often appears in applications, see Sect.4 below.
\vskip 1mm \noindent
\textbf{(iv)} \emph{DL-1,2:} Equations  \eqref{2-bar-0}, and \eqref{3-bar-1} coincide with the original equation \eqref{ODE1}, taken for the $\tau$-independent function $\vf$;
\vskip 1mm \noindent
\textbf{(v)} \emph{DL-2:} Equation \eqref{2-bar-1} represents the linearized version of \eqref{2-bar-0}.
Equation \eqref{2-bar-2} describes the second-order perturbations for \eqref{2-bar-0}, complemented by $\overline{\vF}^{\text{vg}}$.
For a better demonstration of the effects of the vibrogenic force, one can choose $\overline{\vx}_1(s)\equiv 0$ in \eqref{2-bar-0}-\eqref{2-bar-2}, which yields:
\begin{eqnarray}
&&\overline{\vx}_{0ss}=\overline{\vf}_0,\quad \overline{\vx}_{2ss}=(\overline{\vx}_2\cdot\nabla_0)\overline{\vf}_0+\overline{\vF}^{\text{vg}}
\label{2-bar-2a}
\end{eqnarray}
\noindent
It reveals the appearance of $\overline{\vF}^{\text{vg}}$ for linear perturbations, which is promising for various applications.
In particular, it can be important for the studies of long-time evolution, for deviations from symmetry, \emph{etc.} Relevant applications can be found, for example, in astronomy.

\vskip 1mm \noindent
\textbf{(vi)} \emph{DL-1, link to physics:}
The \emph{DL-1} corresponds to $\kappa=1$ which gives $\alpha+\beta=1$, see \eqref{PDE1}. Two physically promising cases are $(\alpha,\beta)=(0,1)$ (when $s=t$ and $\varepsilon=1/\omega$) and $(\alpha,\beta)=(1,0)$ (when $s=t/\omega$ and $\varepsilon=1/\omega^2$).
In both cases, the same $\overline{\vF}^{\text{vg}}$ \eqref{F-vibr} appears in the average equations of zeroth approximation, however the slow time variables are different.

\vskip 1mm \noindent
\textbf{(vii)}
\emph{How to compare DL-1 and DL-2?}
Let us consider solutions with $\beta=0$ in \eqref{PDE1}.
They could be interesting physically, since they  correspond to the dimensionless force of order one.
Then $(\alpha, \beta)=(1,0), s=\varepsilon t$ in \emph{DL-1}, and $(\alpha, \beta)=(0,0), s=t$ in \emph{DL-2}.
The \emph{DL-1} averaged equation \eqref{F-vibr} (where always $\overline{\vf}_0\equiv 0$ by solvability) is:
\begin{eqnarray}\label{2-trivial}
  \overline{\vx}_{0ss}=\overline{\vF}^{\text{vg}}, \quad\text{where}\quad s=\varepsilon t
 \end{eqnarray}\noindent
The \emph{DL-2} averaged equations are  \eqref{appr-22-bar}, \eqref{appr-23-bar},    \eqref{appr-24-sol-F}.
Considering a special case $\overline{\vf}_0\equiv 0$ yields:
 \begin{eqnarray}\label{1-trivial}
  \overline{\vx}_{0ss}=0,\quad  \overline{\vx}_{1ss}=0,\quad  \overline{\vx}_{2ss}=\overline{\vF}^{\text{vg}}, \quad\text{where}\quad s= t
 \end{eqnarray}
Then  a special \emph{DL-2} solution can be chosen as
\begin{eqnarray}\label{1-trivial}
  \overline{\vx}_{0}=0,\quad  \overline{\vx}_{1}=0,\quad  \overline{\vx}_{2ss}=\overline{\vF}^{\text{vg}}
 \end{eqnarray}
One can see that in terms of the original physical time $t$, the equations  can be (informally) rewritten as
 \begin{eqnarray}\label{12-trivial}
    \overline{\vx}_{2tt}=\overline{\vF}^{\text{vg}}\ \text{for \emph{DL-2}}, \quad \text{and}\quad \overline{\vx}_{0tt}=\varepsilon^2\overline{\vF}^{\text{vg}}\ \text{for \emph{DL-1}}
 \end{eqnarray}
 which shows that the main (zeroth) \emph{DL-1} approximation is the same as the second approximation $\varepsilon^2\overline{\vx}_{2}$ of  the special \emph{DL-2} solution \eqref{1-trivial}.
 Making these asymptotic solutions identical to each other violates our strict setting, in which all our functions are of order one. However, an informal renormalization of the involved function allows us to conclude that, in some cases, the solutions \emph{DL-1} represents a special case of \emph{DL-2}.

\vskip 1mm \noindent
\textbf{(viii)} $\kappa\ge 3$: For $\kappa=3$ system  \eqref{3-bar-0}-\eqref{3-bar-4} degenerates.
It appears since all the coefficients in Taylor's series \eqref{Taylor} are calculated at the arbitrary rectilinear motion $\vx_0=\overline{\vx}_0=\va s+\vb$. 
In this situation all the averaged equations have a structure when `an acceleration is equal to a given force', hence the secular behaviour of their solutions is common. For example, one can consider the state of rest $\vx_0=\overline{\vx}_0=0$. Then \eqref{3-bar-1} simplifies to $\overline{\vx}_{1ss}=\overline{\vf}(0)=\const$, which gives the quadratic growth of $\overline{\vx}_1(s)$, \emph{etc.}
Similar degeneration takes place for the averaged equations with any $\kappa\ge 3$.
Therefore, one can consider all these degenerated cases as \emph{useless for applications}. The question of their mathematical validity remains open.

\vskip 1mm \noindent
\textbf{(ix)} \emph{Potential forces:}
For a potential vibrational force $\widetilde{\vf}=\nabla \widetilde{\varphi}$ the vibrogenic force also has a potential:
\begin{eqnarray}\label{Vibro-potent}
&&\overline{\vF}^{\text{vg}}=-\nabla\overline{\Pi}_0,\quad \text{where}\quad\overline{\Pi}_0\equiv\langle(\nabla\widetilde{\varphi}_0^{\tau})^{ 2}\rangle/2
\end{eqnarray}
Hence, for potential forces (like gravitational, magnetostatic,  or electrostatic ones) the studies are simplified.

\section{The Stephenson-Kapitza Pendulum fitted to Distinguished Limits}

The mathematical part of the job is over after all the \emph{DLs} are determined and asymptotic equations/solutions are calculated.
The next step is to choose the values of $\alpha$ and $\beta$ in \eqref{PDE1}, which are suitable for particular physical applications.
Let us consider a mathematical pendulum with a vertically vibrating pivot.
Such a pendulum represents a paradigm of Vibrodynamics and averaging methods,
therefore it is instructive to demonstrate how it can be seen in the \emph{DLs} framework.
The dimensional equation is
\begin{eqnarray}\label{A1}
l\theta_{t^\ast t^\ast}=-(g+Y^\ast_{t^\ast t^\ast})\sin\theta
\end{eqnarray}
where $\theta$ is the angle of pendulum's deviation from the vertical, $g$ is the homogeneous vertical gravity field, $l$ is the length of the pendulum, $Y^\ast(\tau)$ is the periodically changing vertical coordinate of the pivot, $\tau\equiv\omega^\ast t^\ast$, $\omega^\ast$ is the given frequency of the pivot's oscillations.
The use of characteristic scales of length $L=l$ and time $T=\sqrt{l/g}$ leads to the dimensionless equation:
\begin{eqnarray}
\theta_{tt}=-(1+ \omega^2\widetilde{Y}_{\tau\tau})\sin\theta,\quad \widetilde{Y}=\widetilde{Y}(\tau),\ \tau=\omega t\label{A2}
\end{eqnarray}
This equation represents a special one-dimensional (scalar) form of \eqref{ODE1}; in \eqref{A2} the unknown function $\theta(t)$ plays a part of $x(t)$.
If $\widetilde{Y}\sim O(1)$ then \eqref{A2} is out of our consideration due to the main constraint \eqref{smal-tilde}.
Hence, we abolish the requirement $\widetilde{Y}\sim O(1)$ by introducing either of the two constraints for the amplitude of pivot's vibrations
\begin{eqnarray}
&&\text{(a)}\quad \widetilde{Y}=\widetilde{\xi}(\tau)/\omega,\quad \text{(b)}\quad \widetilde{Y}=\widetilde{\xi}(\tau)/\omega^2, \label{V-fin-0}
\end{eqnarray}
where $\widetilde{\xi}(\tau)\sim O(1)$ is a given function.
In the limit $\omega\to\infty$, the case (a) corresponds to the finite velocity and infinite acceleration,
while (b) gives the limit of zero velocity and finite acceleration.
Then the particular forms of \eqref{ODE1} are
\begin{eqnarray}
&&\text{(a)}\quad x_{tt}=-(1+ \omega\widetilde{\xi}_{\tau\tau})\sin x,\quad  \text{(b)}\quad x_{tt}=-(1+ \widetilde{\xi}_{\tau\tau})\sin x,\label{B-fin-1}
\end{eqnarray}
where, to keep similarity with \eqref{ODE1}, the unknown function $\theta$ is replaced with $x$.
Then the only component of force $f(x,\tau)$ appears as
\begin{eqnarray}
&&\text{(a)}\quad \quad x_{tt}= \omega (f(x,\tau)+g(x,\tau)/\omega)=-\omega(\widetilde{\xi}_{\tau\tau}(\tau)+1/\omega)\sin x\label{A-fin-2}\\
&&\text{(b)}\quad  \quad x_{tt}=f(x,\tau)=-(\widetilde{\xi}_{\tau\tau}(\tau)+1)\sin x \label{B-fin-2}
\end{eqnarray}
The minor difference between \eqref{ODE1} and \eqref{A-fin-2} is:  instead of the function $f$  in \eqref{ODE1}, we use a sightly more general function $f+ g/\omega$, where both functions $f$ and $g$ are of order one.
As one can see, (a) belongs to \emph{DL-1} with $(\alpha, \beta)=(0, 1)$, and (b) belongs to \emph{DL-2} with $(\alpha, \beta)=(1, 0)$.
In the case (a), the averaged \emph{DL-1} equation \eqref{1-bar-0} follows automatically (we return notations back from $x$ to $\theta$):
\begin{eqnarray}
&&\overline{\theta}_{0ss}=-\sin\overline{\theta}_0+\overline{F}^{\text{vg}},\quad \overline{F}^{\text{vg}}= -\frac{1}{2}  \langle\widetilde{\xi}^2_\tau\rangle\sin(2\overline{\theta}_0)\label{A-fin-av}
\end{eqnarray}
The averaged equations for (b) are also straightforward, we do not write them here.
The \emph{DL-1} equation \eqref{A-fin-av} leads to the conservation of `energy'
\begin{eqnarray}
&&E_0=K_0+\Pi_0=\const,\ K_0\equiv(\overline{\theta}_{0s})^2/2,\   \Pi_0\equiv-\cos\overline{\theta}_0-   (\langle\widetilde{\xi}^2_\tau\rangle/4)\cos(2\overline{\theta}_0)\label{A-energy-av}
\end{eqnarray}
which is always used for the demonstration of \emph{the upside down pendulum or inverted pendulum}.
The upside down equilibrium at $\overline{\theta}_0=\pi$ appears due to the vibrogenic term in the potential energy $\Pi_0(\overline{\theta}_0)$, this term agrees with \eqref{Vibro-potent}.
A physically minded reader may be surprised here that \emph{Nature} has chosen the limit \emph{DL-1}, which corresponds to infinite acceleration.

\section{Conclusions and Discussion}

\emph{A Practical Algorithm:} Our consideration leads to a simple and highly practical algorithm of building averaged equations for \eqref{ODE}-\eqref{PDE}. 
The only two productive options are \emph{DL-1} with the averaged equation \eqref{1-bar-0} and \emph{DL-2} with the averaged equations \eqref{2-bar-0}-\eqref{2-bar-2}.
The latter, for a better physical exposition, can be replaced by \eqref{2-bar-2a}.
The choice between these two options is determined by the structure of the main approximation of the force $\vf_0\equiv\vf(\vx_0,\tau)$.
Under the  constraint  $\eqref{smal-tilde}$, it takes a form $\vf_0=\vf(\overline{\vx}_0(s),\tau)$.
If the average $\overline{\vf}_0\equiv\langle\vf(\overline{\vx}_0(s),\tau)\rangle=0$, then we have the case of \emph{DL-1}.
If $\overline{\vf}_0\neq 0$, then we deal with \emph{DL-2}.
The value of $\overline{\vf}_0$ is easy to calculate, since $\overline{\vx}_0(s)$ is treated as a constant.
After the choice between \emph{DL-1} and \emph{DL-2} has been done, the only remaining job is to calculate $\overline{\vF}^{\text{vg}}$ \eqref{F-vibr}, and then go for a particular physical application.


\emph{Second Order ODEs \emph{versus} First Order ODEs and PDEs:} Mathematically, a system of the first order ODEs is more general, than a second order equation. However, the obtaining of the second-order equation from a system of first-order equations requires a well known degeneration of the latter. Therefore the second-order ODEs are worth to consider separately. Very similar ideas can be used for PDEs, see \emph{e.g.} \cite{Vladimirov, Vladimirov1}.


\section{Appendix: Calculations for \emph{DL-1} and \emph{DL-2}}

\underline{Calculations for the \emph{DL-1} ($\kappa=1$)}:
Equation \eqref{PDE1} takes the form:
\begin{eqnarray}
&&{\vx}_{\tau\tau}+ 2\varepsilon\vx_{\tau s}+\varepsilon^2 \vx_{ss} =\varepsilon\vf(\vx,\tau)\label{PDE-1a}
\end{eqnarray}
The substitution of \eqref{series}, \eqref{Taylor} into \eqref{PDE-1a} produces the successive approximations:
\vskip 1mm

\emph{The zero-order equation}  (\ref{PDE-1a})  is $\vx_{0\tau\tau}=0$.
The substitution of $\vx_{0}=\overline{\vx}_0(s)+\widetilde{\vx}_0(s,\tau)$ gives $\widetilde{\vx}_{0\tau\tau}= 0$.
 Integrating it twice yields $\widetilde{\vx}_0=\vh_1(s)\tau+\vh_2(s)$, where the arbitrary functions $\vh_1(s)$ and $\vh_2(s)$ vanish due to the $\tau$-periodicity and zero average of $\widetilde{\vx}_0$.
Hence, the unique solution is $\widetilde{\vx}_0\equiv 0$ and the results are:
\begin{eqnarray}
&&\widetilde{\vx}_0(s,\tau)\equiv 0;\quad \overline{\vx}_0(s) \ \text{is not defined}\label{appr-sol-10}
\end{eqnarray}
\emph{The first-order equation} (\ref{PDE-1a})  is $\vx_{1\tau\tau}+2\vx_{0\tau s}=\vf_0$.
Its averaging leads to the constraint $\overline{\vf}_0\equiv 0$, whilst the tilde-part is $\widetilde{\vx}_{1\tau\tau}=\widetilde{\vf}_0$.
Hence, we obtain:
\begin{eqnarray}
&&\widetilde{\vx}_1(s,\tau)=\widetilde{\vf}_0^{\tau\tau};\quad \overline{\vx}_1(s) \ \text{is not defined}\label{appr-sol-11}
\end{eqnarray}
\emph{The second-order equation} (\ref{PDE-1a}) is
\begin{eqnarray}
&&\vx_{2\tau\tau}+2\vx_{1\tau s}+\vx_{0ss}=(\vx_1\cdot\nabla_0)\vf_0\label{appr-13}
\end{eqnarray}
Its bar-part appears as:
$\overline{\vx}_{0ss}=\langle(\widetilde{{\vx}}_1\cdot\nabla_0)\widetilde{{\vf}}_0\rangle$;
the substitution of \eqref{appr-sol-11} into this expression and integration by parts lead to \eqref{1-bar-0} and \eqref{F-vibr}:
\begin{eqnarray}
&&\overline{\vx}_{0ss}=\overline{\vF}^{\text{vg}},\quad\overline{\vF}^{\text{vg}}\equiv -\langle(\widetilde{\vf}_{0}^\tau\cdot\nabla_0)\widetilde{\vf}_{0}^\tau\rangle\label{DL-1-eqn}
\end{eqnarray}
Since the \emph{vibrogenic force} $\overline{\vF}^{\text{vg}}$ appears in the zeroth approximation, we do not present the calculations for any further steps. We claim  that any further approximation can be calculated, including its average and oscillatory parts.
\vskip 3mm

\underline{Calculations for the \emph{DL-2} ($\kappa=2$)}:
Equation \eqref{PDE1} takes the form:
\begin{eqnarray}
&&{\vx}_{\tau\tau}+ 2\varepsilon\vx_{\tau s}+\varepsilon^2 \vx_{ss} =\varepsilon^2\vf(\vx,\tau)\label{PDE-2}
\end{eqnarray}
The substitution of \eqref{series}, \eqref{Taylor} into \eqref{PDE-2} produces the successive approximations:
\vskip 1mm
\noindent
\emph{The zero-order equation}  (\ref{PDE-2})  is again $\vx_{0\tau\tau}=0$.
Hence, the results are the same as  \eqref{appr-sol-10}:
\begin{eqnarray}
&&\widetilde{\vx}_0(s,\tau)\equiv 0;\quad \overline{\vx}_0(s) \ \text{is not defined}.\label{appr-sol20}
\end{eqnarray}
\emph{The first-order equation} (\ref{PDE-2})  is $\vx_{1\tau\tau}+2\vx_{0\tau s}=0$.
It leads to
\begin{eqnarray}
&&\widetilde{\vx}_1(s,\tau)\equiv 0;\quad \overline{\vx}_1(s) \ \text{is not defined}.\label{appr-sol-21}
\end{eqnarray}
\emph{The second-order equation} (\ref{PDE-2}) is
\begin{eqnarray}
&&\vx_{2\tau\tau}+2\vx_{1\tau s}+\overline{\vx}_{0ss}= \overline{\vf}_0+\widetilde{\vf}_0,\label{appr-22}
\end{eqnarray}
which produces the bar-part \eqref{3-bar-0}
\begin{eqnarray}
\overline{\vx}_{0ss}=\overline{\vf}_0,\label{appr-22-bar}
\end{eqnarray}
whilst the tilde-part leads to
\begin{eqnarray}\label{appr-22-tilde}
&&\widetilde{\vx}_{2\tau\tau}=\widetilde{\vf}_0, \quad \widetilde{\vx}_2=\widetilde{\vf}_0^{\tau\tau}.
\end{eqnarray}
\emph{The third-order equation} (\ref{PDE-2})
\begin{eqnarray}
&&\vx_{3\tau\tau}+2\vx_{2\tau s}+\vx_{1ss}=(\vx_1\cdot\nabla_0)\vf_0\label{appr-23}
\end{eqnarray}
Its averaging with the use of \eqref{appr-sol-21} produces \eqref{2-bar-1}:
\begin{eqnarray}
\overline{\vx}_{1ss}=(\overline{\vx}_1\cdot\nabla_0)\overline{\vf}_0
\label{appr-23-bar}
\end{eqnarray}
\emph{The forth-order equation} (\ref{PDE-2})
\begin{eqnarray}
&&\vx_{4\tau\tau}+2\vx_{3\tau s}+\vx_{2ss}=(\vx_2\cdot\nabla_0)\vf_0+\frac{1}{2}x_{1i}x_{1k}\vf_{0,ik} \label{appr-24}
\end{eqnarray}
produces the bar-part
\begin{eqnarray}
\overline{\vx}_{2ss}=(\overline{\vx}_2\cdot\nabla_0)\overline{\vf}_0+\langle(\widetilde{\vx}_2\cdot\nabla_0)\widetilde{\vf}_0\rangle+
\frac{1}{2}\overline{x}_{1i}\overline{x}_{1k}\overline{\vf}_{0,ik}
\label{appr-24-sol}
\end{eqnarray}
The substitution of $\widetilde{\vx}_2$ from \eqref{appr-22-tilde} and integration by parts produce \eqref{2-bar-2}
\begin{eqnarray}
\overline{\vx}_{2ss}=(\overline{\vx}_2\cdot\nabla_0)\overline{\vf}_0+
\frac{1}{2}\overline{x}_{1i}\overline{x}_{1k} \overline{\vf}_{0,ik}+\overline{\vF}^{\text{vg}}
\label{appr-24-sol-F}
\end{eqnarray}
where the \emph{vibrogenic force} $\overline{\vF}^{\text{vg}}$ is again given by \eqref{F-vibr}.
Since its appearance, we do not present the calculations of further approximations; we claim here that \emph{DL-2} also produces a valid asymptotic procedure.

\begin{acknowledgments}
\emph{Acknowledgements:} The paper is devoted to the memory of Prof. A.D.D. Craik, FRSE, who read this manuscript and made useful critical comments. A special thank to Prof. M.R.E. Proctor, FRS, for a productive discussion, and to Mr. A. A. Aldrick for help with the manuscript. This research is partially supported by the grant IG/SCI/DOMS/ 18/16 from SQU, Oman.
\end{acknowledgments}

\end{document}